\documentclass[twocolumn,prb,showpacs,floatfix]{revtex4-1}
\usepackage{graphicx}
\usepackage{amsmath}
\usepackage{subfigure}
\usepackage{hyperref}
\usepackage{tikz}
\usepackage{xcolor}
\usepackage{float}
\usepackage{gensymb}

\usetikzlibrary{decorations.pathmorphing,decorations.markings,arrows}
\begin{document}
\tikzset{->-/.style={decoration={
  markings,
  mark=at position #1 with {\arrow{>}}},postaction={decorate}}}

\title{Topological phases in Iridium oxide superlattices: quantized anomalous charge or valley Hall insulators}
\author{Yige Chen$^1$ and Hae-Young Kee$^{1,2,}$}
\email{hykee@physics.utoronto.ca}
\affiliation{Department of Physics, University of Toronto, Ontario M5S 1A7 Canada}
\affiliation{Canadian Institute for Advanced Research, CIFAR Program in Quantum Materials, Toronto, ON M5G 1Z8, Canada}
\date{October 28, 2014}
\begin{abstract}
We study topological phases in orthorhombic perovskite Iridium (Ir)  oxide superlattices grown along the $[001]$ crystallographic axis. 
Bilayer Ir oxide superlattices display topological magnetic insulators exhibiting quantized anomalous Hall effects due to strong spin-orbit coupling of Ir 5d-orbitals and electronic correlation effects.
We also find a valley Hall insulator with counter-propagating edge currents from two different valleys 
and a topological crystalline insulator with edge states protected by the crystal lattice symmetry based on stacking of two layers. 
In a single layer superlattice, a topological insulator can be realized, when a strain field is applied to break the symmetry of a glide plane protecting the Dirac points.
It turns into a  topological magnetic insulator in presence of magnetic ordering and/or in-plane magnetic field.
We discuss essential ingredients for these topological phases and experimental signatures to test our theoretical proposals.
\end{abstract}
\maketitle

\section{Introduction}
Considerable attention has recently been devoted to the study of non-trivial physics arising from strong spin-orbit coupling (SOC). 
Such studies were initiated by theoretical proposals of topological insulators with conducting surface states protected by time reversal (TR) symmetry~\cite{KanePRL05,BernevigPRL06,ShengPRL06,BernevigScience06,FuPRB07,FuPRL07,MoorePRB07,Roy06,KanePW11},
which was then experimentally confirmed  in two-dimensional (2D) HgTe/Hg$_{1-x}$Cd$_x$Te quantum wells~\cite{KonigScience07}
and  indirectly by angle resolved photoemission spectroscopy (ARPES) in three dimensional (3D) systems such as
Bi$_{1-x}$Sb$_x$~\cite{HsiehNature08,HsiehScience09}, Bi$_2$Se$_3$~\cite{XiaNP09,HorPRB09}, Bi$_2$Te$_3$ and  Sb$_2$Te$_3$~\cite{HsiehPRL09,HasanRMP10,QiRMP11}. 
Since then, a variety of topological phases have been theoretically suggested. These include topological crystalline insulators  with surface states protected by crystal lattice symmetry~\cite{Fu11, Hsieh12, Xu12, Dziawa12, Tanaka12},  Weyl semimetals with chiral fermions~\cite{Burkov11, Wan11, Volovik07, Murakami07, Yang11}, and topological magnetic insulators with quantized anomalous Hall (QAH)  effects~\cite{Yu10, Liu08, Xu11, Chang13}.  Furthermore, strongly interacting systems could provide a new avenue to explore more exotic phases such as  topological Mott insulators and fractional Chern insulators~\cite{Sondhi13,Krempa14}.


While the number of topological phases proposed in theory is still growing, experimental confirmations
are limited to the systems of groups IV-VI elements. Why have such topological phases not been detected in other abundant materials such as oxides ?
In particular, transition metal oxides exhibit various collective phenomena stemming from strong electronic correlations, and this has led to tremendous interest and effort in growing
 oxide films to discover new functionalities. However, this effort has so far been focus mainly on 3d- and 4d-orbital systems with weak
or moderate SOC, and little attention has been paid to 5d-orbital systems with strong SOC until recently. 


Among 5d-orbital systems, Ir oxides named Iridates have provided an excellent playground to study the combined effects of
SOC and electron correlations. 
Depending on  the underlying lattice structure,
Iridates have offered a rich phase diagram~\cite{Krempa14}. Despite different phases,
a common ingredient is the $J_{\textrm{eff}}=\frac{1}{2}$ description due to strong atomic SOC is a good starting point in building microscopic Hamiltonians.
%
%
Using $J_{\rm eff}=\frac{1}{2}$ wavefunction, a topological insulator was proposed in 3D perovskite Iridates.~\cite{CarterPRB12}
It was found that bulk SrIrO$_3$ with P$_{\rm bnm}$ structure exhibits a crystal-symmetry-protected nodal line which becomes a 3D nodal point when the mirror symmetry along the c-axis is broken. It becomes a topological insulator with large mirror symmetry breaking term.~\cite{CarterPRB12}
A successful growing of Ir oxide superlattice, [(SrIrO$_3$)$_{n}$,SrTiO$_3$] where the integer ${n}$ controls the number of Ir oxide layers
using pulsed laser deposition (PLD) technique has been also reported.~\cite{Matsuno14}
It has demonstrated how a spin-orbit magnetic insulator arises by tuning the number of SrIrO$_3$ layers.
%

Given that SrIrO$_3$ with P$_{\rm bnm}$ structure
possesses a crystal-symmetry-protected nodal line, it is possible to design other topological phases by employing the current experimental  techniques.
While a topological insulator was proposed in an effective honeycomb bilayer by fabricating [111] superlattice structure from perovskite oxides ~\cite{XiaoNC11},
atomically controlled [111] superlattice of perovksite oxides is known difficult to be fabricated.
 On the other hand, Ir oxide superlattice along the [001] axis
has been successfully made by J. Matsuno et al. ~\cite{Matsuno14} as stated above. 
In this paper, we show how to realize topological phases in Ir oxide superlattices grown along the [001] axis; [(SrIrO$_3$)$_{n}$, (AMO$_3$)$_{n^\prime}$] for integer $n^\prime$ and
${n}=1$ or 2 where AMO$_3$ is a band insulator with a closed shell transition metal $M^{4+}$ and
an alkaline earth metal $A^{2+}$. To realize topological phases, one has to retain oxygen octahedra rotation {\it and} tilting which is necessary to generate a  Rashba-like SOC in
$J_{\rm eff}=\frac{1}{2}$ basis. Thus AMO$_3$ should have the orthorhombic P$_{\rm bnm}$ structure such as CaTiO$_3$,  SrZrO$_3$, or SrHfO$_3$ instead of SrTiO$_3$ with tetragonal structure. 
The topological states realized in these superlattices include topological magnetic insulators with QAH effects, non-trivial valley insulators, topological
insulators with TR symmetry, and topological crystalline insulators.

This paper is organized as follows.
In Sec. 2, we show how a 2D topological insulator can be made in an Ir oxide single layer system. When oxygen octahedron is rotated and titled away from
c-axis, there are two 2D Dirac points similar to the honeycomb lattice~\cite{KanePRL05-2}. These 2D Dirac points are protected by
the b-glide symmetry. 
Breaking this b-glide symmetry generates a 2D topological insulator, and
furthermore in the presence of a magnetic ordering and/or  in-plane magnetic field, the system
becomes a topological magnetic insulator. This could be confirmed by quantized Hall conductance in Hall measurement. 
In Sec. 3, we propose two different types of bilayer Ir oxides. Depending on the layer stacking,
one becomes a topological magnetic insulator for any small magnetic field that breaks the b-glide symmetry.
The other case possesses various topological phases including topological crystalline, topological magnetic, and mirror valley insulators.
In each section, we offer a schematic crystal structure of Ir oxide superlattices and physical origins of such topological phases based on symmetry of lattice and TR.
We summarize our findings in the last section.

\begin{figure}[t]
\centering
\includegraphics[width=8.5cm]{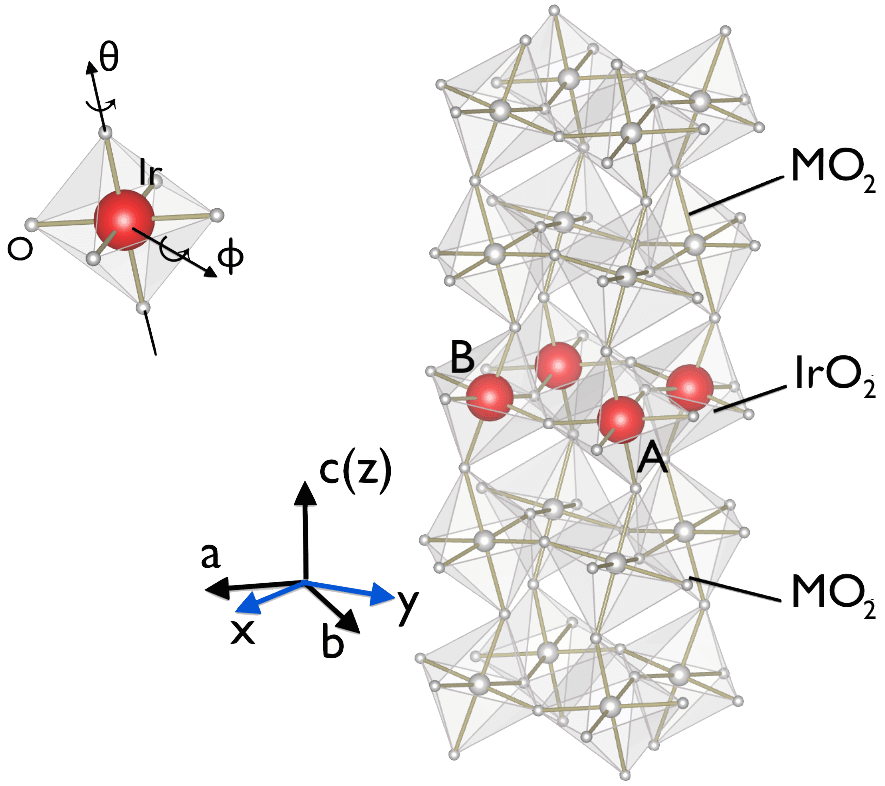}
\caption{(color online) (left) IrO$_6$ octahedron with the rotation $\theta$ along c-axis and titling  $\phi$ along local $(110)$ axis. (right)  Single layer Ir oxide superlattice structure.  IrO$_2$ layer contains two different sites denoted by A and B representing different rotations and tiltings, $(\theta,\phi)$ and $(-\theta, -\phi)$ oxygen octahedra, and it is grown
on a band insulator AMO$_3$ with P$_{\rm bnm}$ structure. The primitive lattice vectors are $\vec{a}=(\hat{x}-\hat{y})/2$ and $\vec{b}=(\hat{x}+\hat{y})/2$.}
\label{fig:slayer}
\end{figure}

\section{Single-Layer Iridates}
\subsection{Model Hamiltonian and Dirac fermion}
In bulk samples AMO$_3$ with P$_{\rm bnm}$ structure, each M atom surrounded with six O atoms forms an octahedron. 
This octahedron is rotated by an angle $\theta$ around the c-axis and tilted by an angle $\phi$ around the local (110) direction as shown in Fig. ~\ref{fig:slayer}. The rotation and tilting angles alternate between two neighboring  IrO$_6$ octahedra in the plane and between adjacent layers making four M atoms in a unit cell. 
To engineer a single-layer Ir oxide, IrO$_2$  layer is grown from AMO$_3$ as shown in Fig. ~\ref{fig:slayer}.
 $x$- and $y$-direction are rotated by 45 degree from the crystal $a$- and $b$-axis for convenience.
As we state above, the alternating rotation and tilting of neighboring IrO$_6$ is crucial to realize topological phases for the following reason.
The relatively strong SOC of Ir atoms splits t$_{\rm 2g}$ states into $J_{\rm eff}=\frac{1}{2}$ and $J_{\rm eff}=\frac{3}{2}$,
and Ir$^{4+}$ ionic configuration 
leading to the valence of $5d^{5}$  makes 
these iridates to be a half-filled $J_{\rm eff}=\frac{1}{2}$ band. Even though the tetragonal distortion of IrO$_6$ octahedra may affect the validity of the $J_{\rm eff}=\frac{1}{2}$ description in reality, the tetragonal crystal field splitting is small compare to the SOC of iridium~\cite{JinPRB09,AritaPRL12}. Thus, $J_{\rm eff}=\frac{1}{2}$ states are well separated from $J_{\rm eff}=\frac{3}{2}$ states, which makes $J_{\rm eff}=\frac{1}{2}$ picture still adequate to describe the physics near the Fermi energy.    Note that $J_{\rm eff} =\frac{1}{2} $ consists of
$| J_z = \pm \frac{1}{2} \rangle = \frac{1}{\sqrt{3}} \left(  |d_{xy,s} \rangle \pm |d_{yz,-s} \rangle + i |d_{xz,-s} \rangle \right)$ 
where  $\pm s$ represents spin-1/2  up and down states~\cite{footnoteaxis}, respectively.
In the presence of  the alternating tilting and rotation between neighboring sites, 
a hopping integral between $d_{xy,s}$ and $d_{xz/yz,s}$ orbitals becomes finite.
Since $d_{xy,s}$ and $d_{xz/yz,s}$ belong to different spin states of $|J_z\rangle$,
this hopping involves $|J_z= \frac{1}{2} \rangle$ and $|J_z =-\frac{1}{2} \rangle$ states which then generates a spin-flip Rashba-like term.

For a single layer of IrO$_2$, there are two sites due to different rotation ($\theta$) and tilting angle ($\phi$) between nearest-neighbor sites. We denote these Ir sites by
 A and B indicating different oxygen environments as shown in Fig.~\ref{fig:slayer}. It has a rectangle structure associated with a glide symmetry plane which corresponds to 
the invariance under a 1/2 translation along a certain direction, and reflection afterwards. In this lattice, it is along $b$-axis and thus named the b-glide.
The effect of this glide plane on $t_{2g}$ orbitals is to interchange $d_{yz}$ with $d_{xz}$ orbital and to exchange A with B site.
Introducing the Pauli matrices $\vec{\tau}$ and $\vec{\sigma}$ for
the sublattice $A$ and $B$, and $J_{\rm eff}=1/2$ pseudospin, respectively, this b-glide symmetry plane is 
 expressed as, 
\begin{eqnarray}
\hat{\Pi}_b=\frac{i}{\sqrt{2}}(\sigma_x-\sigma_y)\tau_x \hat{k}_{bg},
\label{eq:a1}
\end{eqnarray}
where $\hat{k}_{bg}$ is the operator acting on crystal momentum space as $\hat{k}_{bg} : (k_x,k_y) \rightarrow (k_y, k_x)$.~\cite{CarterPRB12}

\begin{figure*}
\subfigure[$\phi=0$ with b-glide symmetry]{
\includegraphics[width=5.5cm]{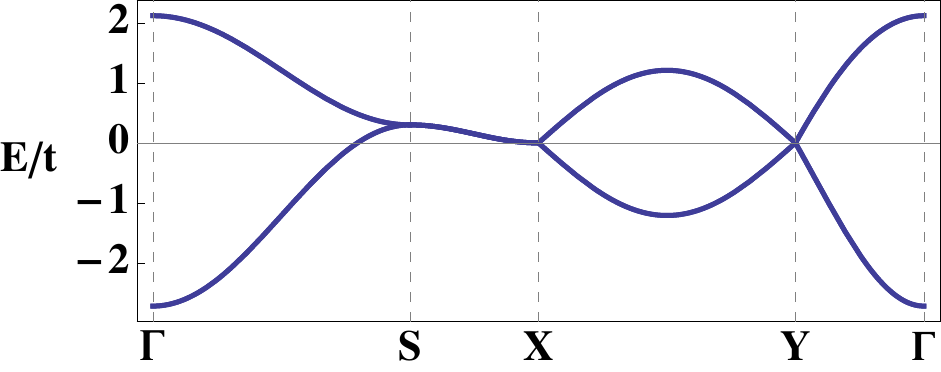}
\label{fig:notiltrot}
}
\subfigure[Finite $\phi$ with b-glide symmetry]{
\includegraphics[width=5.5cm]{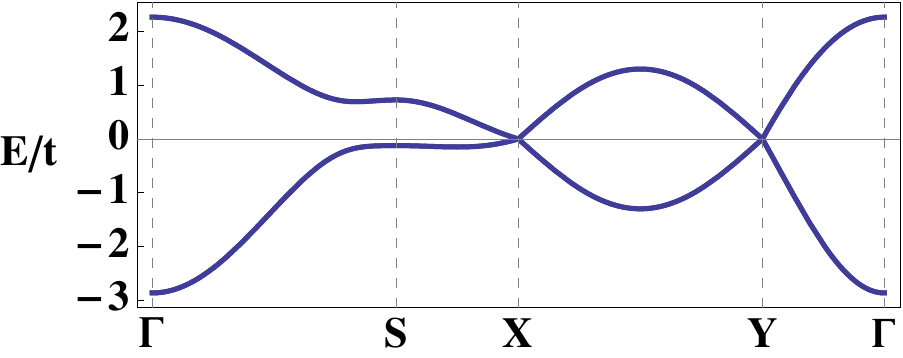}
\label{fig:bglidehas}
}
\subfigure[Finite $\phi$ without b-glide symmetry]{
\includegraphics[width=5.5cm]{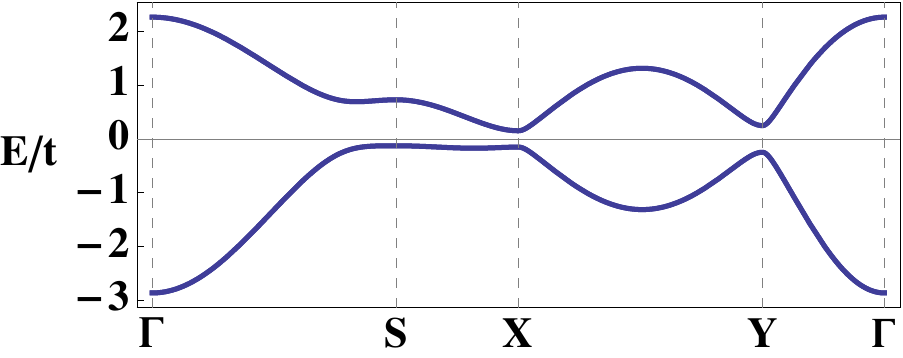}
\label{fig:bglideno}
}
\caption{(color online) Band dispersion of single layer Ir oxide (a)  without tilting $\phi$. It shows four fold degeneracy along $S=(\pi,0) \rightarrow X=(\frac{\pi}{2},-\frac{\pi}{2})$ direction. 
(b) Finite rotation and tilting leaves two Dirac points at $X$ and $Y=(\frac{\pi}{2},\frac{\pi}{2})$. (c) When the b-glide symmetry is broken, Dirac point acquires a finite gap at $X$ and $Y$
points. The set of $(\theta,\phi)$ for both (b) and (c) is $(7^{\circ},19^{\circ})$. }
\label{fig:singleband}
\end{figure*}

A tight-binding model can be constructed from $J_{\textrm{eff}}=1/2$ bands with the basis $(A \uparrow, B \uparrow, A \downarrow, B \downarrow)$ where A and B denote two different Ir sites in the unit cell as discussed above, and $(\uparrow, \downarrow)$ represents $J_z=\pm \frac{1}{2}$. 
Taking into account nearest and next-nearest hoppings, the Hamiltonian is given by
\begin{eqnarray}
H_0({\bf{k}})
&=&\epsilon_0({\bf{k}})\tau_x + \epsilon^{\prime}({\bf{k}}) {\bf I} \nonumber\\
&+&\epsilon_{1d}({\bf{k}})\sigma_z\tau_y+\epsilon_y({\bf{k}})\sigma_y\tau_y+\epsilon_x({\bf{k}})\sigma_x\tau_y,
\label{eq:a3}
\end{eqnarray}
where
\begin{eqnarray}
\epsilon_{0/1d}({\bf{k}})&=&2t_{0/1d}(\cos(k_x)+\cos(k_y)),\nonumber\\
\epsilon_{y/x} ({\bf{k}})&=& t_1\cos(k_{x/y})+t_{2}\cos(k_{y/x}),\nonumber\\
\epsilon^\prime({\bf k})&=&t^\prime\cos(k_x)\cos(k_y).
\label{eq:a4}
\end{eqnarray}
Here $t_0$ is the nearest neighbor (NN) intra-orbital hopping and $t_{1d}$ is the NN hopping between $d_{yz}$ and $d_{xz}$ orbitals. 
$t^\prime$ is the next-nearest neighbor (NNN) intra-orbital hopping. 
$t_1$ and $t_2$ are the NN hopping from $d_{yz}$ and $d_{xz}$ orbitals to $d_{xy}$ orbital, respectively. $t_{1d}$, $t_1$ and $t_2$ vanish without the rotation and tilting of octahedra. The hopping parameters are obtained based on Slater-Koster method~\cite{Slater54} and the parameters
are functions of $\theta$ and $\phi$. For example, they are given by $(t^\prime, t_0, t_{1d}, t_{1}, t_{2})/t=(-0.3,-0.6,-0.15,0.15,0.45)$ when $(\theta,\phi)\approx(7^{\circ},19^{\circ})$, where 
$t$ is the $\pi$-bonding between d-orbitals $t_{dd\pi}$, and we set $t_{dd\pi}:t_{dd\sigma}: t_{dd\delta} = 1:\frac{3}{2}:\frac{1}{4}$. Note that the tight-binding parameters are fully determined by a set of $(\theta,\phi)$. Different values of $(\theta,\phi)$ will simply modify the detailed shape of the band dispersion. Thus, by tunning the magnitude of $(\theta,\phi)$, it is possible to have the electron and hole pockets near Fermi energy. However, the topological feature of the band structure (characterized by the Chern numbers) remains intact. This particular choice of $(\theta,\phi)$ is made to avoid the electron and hole pockets at $\epsilon_F$ but topological properties do not depend on the choice of $(\theta,\phi)$.

The band structure  is shown in  Fig.~\ref{fig:singleband}. Without the tilting angle $\phi$,  two bands are degenerate along  $X=(\frac{\pi}{2},-\frac{\pi}{2})$ to $S=(\pi,0)$ as shown in Fig~\ref{fig:notiltrot}.
However, when both rotation and tilting of octahedra are present, this degeneracy is broken,
and there are two Dirac points at $X$ and $Y$ protected by the b-glide symmetry as shown in Fig.~\ref{fig:bglidehas}. 
The Dirac point may appear below the Fermi energy $\epsilon_F$ when the tilting angle $\phi$ is not significant ($\phi < 17^{\circ}$). 
Indirect hopping via the oxygens can change the strength of hopping parameters as well, but the topological nature of phases described here is not altered
by such quantitative changes. 
When the b-glide symmetry is broken, for example by a strain field along x-direction, these Dirac points are gapped as shown in
Fig. ~\ref{fig:bglideno}. In the following subsection, we discuss the topological nature of this insulator by providing the corresponding Chern numbers
and edge state analysis.

\begin{figure*}[t]
\subfigure[TI]{
\includegraphics[width=7.9cm]{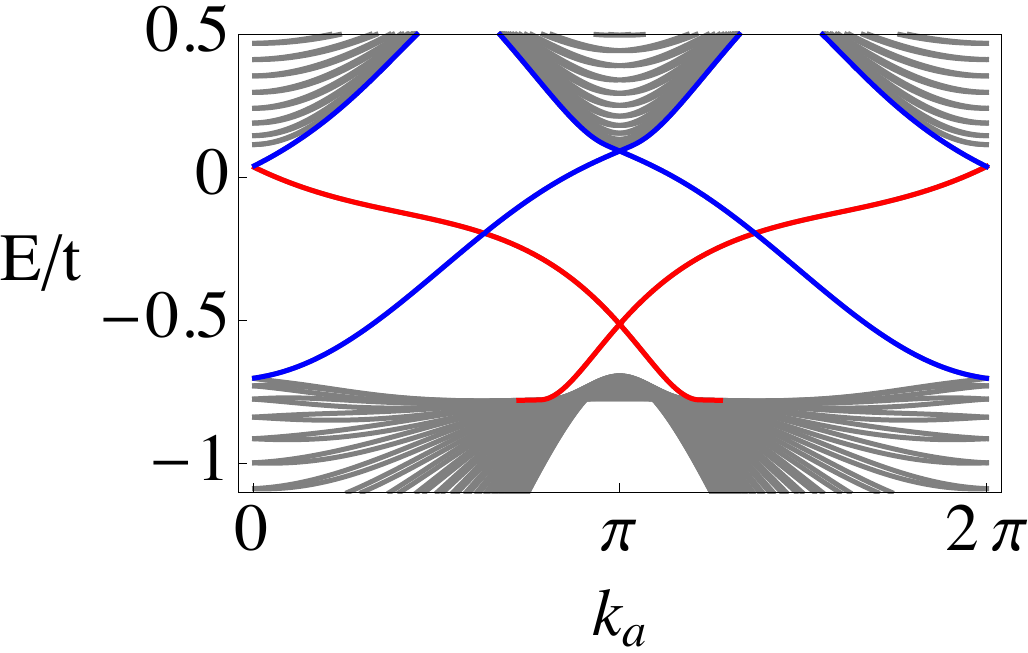}
\label{fig:a1TI}
}
\subfigure[QAHI]{
\includegraphics[width=8cm]{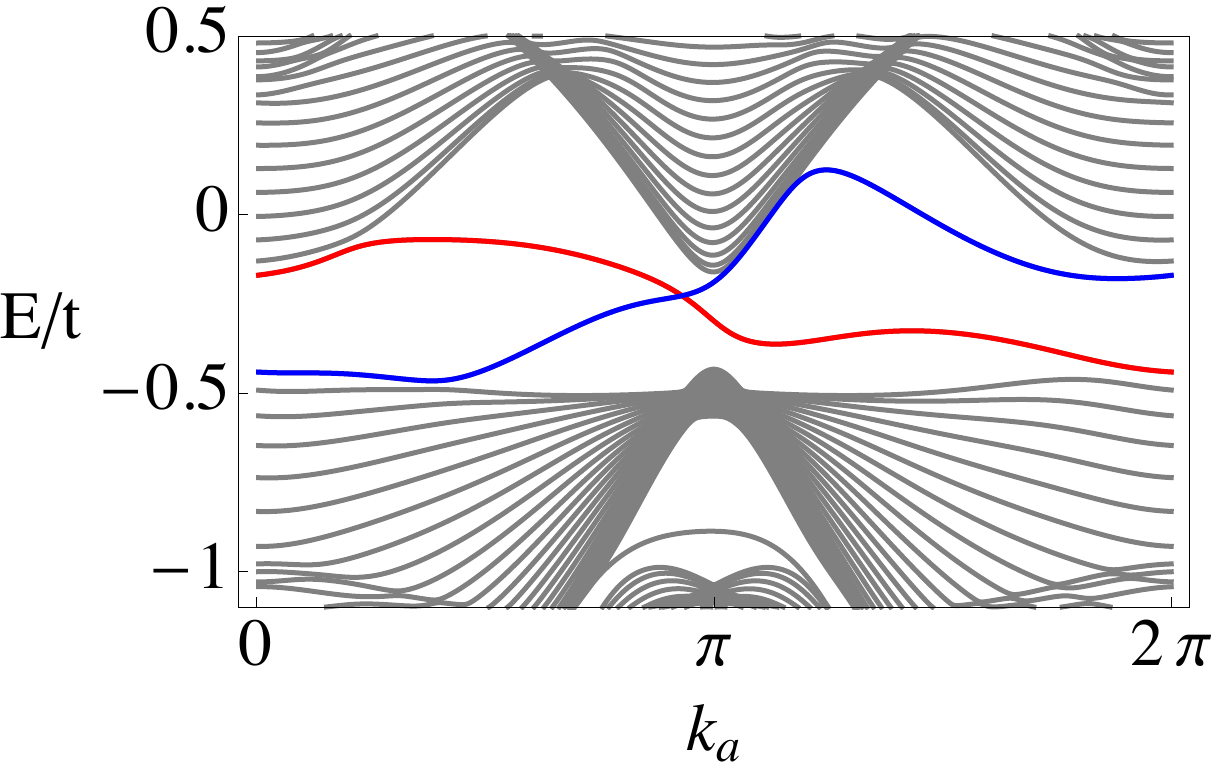}
\label{fig:a1MTI}
}
\caption{(color online) Edge state calculation of (a) topological insulator (TI) shown in  Fig.~\ref{fig:bglideno} and (b) quantized anomalous Hall insulator (QAHI) when TR is broken due to a non-collinear magnetic ordering or an in-plane magnetic field. Grey lines represent bulk state and red (blue) lines denotes edge state at $L=0$ ($L=N$) plotted along $k_a=k_x-k_y-\pi$. The parameter set is the same with the band dispersion in Fig.~\ref{fig:bglideno}. The two gapless edge modes at $L=0$/$L=N$ (red/blue) crossing at 1D TR invariant momentum indicates the system belongs to 2D TI. After breaking the TR, only one gapless edge state left propagating along the boundary.}
\label{fig:a1}
\end{figure*}

\subsection{Topological Insulator and quantized anomalous Hall effects}
Since the Dirac points are protected by the b-glide symmetry, any small perturbation that breaks the b-glide symmetry opens a gap at
these two Dirac points. The b-glide operator is given by Eq.~(\ref{eq:a1}), and thus a small strain along x (or y)-direction is sufficient to break the b-glide symmetry. 
Such a broken b-glide symmetry term allows additional NNN and third NN hoppings as follows.
\begin{equation}
\begin{split}
\epsilon_{2n}({\bf{k}})&=(t_{2n}\cos(k_x+k_y)+t^\prime_{2n}\cos(k_x-k_y))\tau_z,\\
\epsilon_{3n}({\bf{k}})&=2t_{3n}(\cos(2k_x)-\beta \cos(2 k_y))\tau_z,
\end{split}
\label{eq:a5}
\end{equation}
where $t_{2n}$ and $t_{3n}$ are the NNN intra-orbital hopping. $t_{3n}$ is the third NN intra-orbital hopping, and $\beta$ is the parameter to measure the strength of a broken b-glide term. 
The tight-binding parameters  $(t_{2n},t^\prime_{2n},t_{3n})=(0.098,-0.1,0.06)$ obtained by Slater-Koster using the same set of angles ($\theta$,$\phi$) as above, and with $\beta=0.6$ the band dispersion is shown in  Fig.~\ref{fig:bglideno}.

The non-trivial topology behind the gapped Dirac point can be revealed through the following edge state calculation. The slab computation has been performed in a zigzag slab geometry periodic along $\vec{b}=\frac{\hat{x}+\hat{y}}{2}$, while it has an open boundary along $\vec{a}=\frac{\hat{x}-\hat{y}}{2}$; Along $\vec{a}$ direction, one end terminates at atom A and the other side ends with atom B. When TR symmetry is not broken, the system shows gapless edge modes propagating from valence band to conduction band 
as shown in Fig.~\ref{fig:a1TI}. These two gapless edge states cross at a time reversal invariant momentum (TRIM) point indicating their protection by the TR symmetry. 
As long as TR symmetry is present, the degeneracy can not be lifted by disorders or weak interactions. Indeed, we have checked that the edge states are robust, even in the presence
of a random sublattice potential.
$Z_2$ index is another way to confirm the topological insulator. It is straightforward to compute the eigenvalues of the inversion operator~\cite{Fu07}. 
The result shows that $Z_2$ index $=1$ consistent with the edge state calculation.


Another effect of strong SOC in Iridates is an amplification of electronic correlation leading to a spin-orbit Mott insulator. 
The relevant bandwidth $W$ is $J_{\rm eff} =\frac{1}{2}$ band rather than the full t$_{\rm 2g}$ band due to the SOC, and thus the ratio of Hubbard interaction $U$
and the bandwidth $W$ is magnified in Iridates.~\cite{Moon08,KimSci09}  
In order to understand the magnetic ordering pattern, let us consider the Hubbard model with tight-binding Hamiltonian of Eq.~\ref{eq:a3} where $\epsilon_{1d}({\bf k})$ and $\epsilon_{y/x}({\bf k})$ contain pseudospin dependent terms. This NN Hamiltonian can be expressed as
\begin{eqnarray}
H_0=&&\sum_{\langle i.j \rangle}\{t_0 c^{\dagger}_{i,A,\sigma}c_{j,B,\sigma} + i c^{\dagger}_{i,A,\alpha}(\vec{v}\cdot\vec{\sigma})_{\alpha\beta} c_{j,B,\beta}\}+ {\rm h.c.}\nonumber\; ,
\label{eq:r1}
\end{eqnarray}
where $\vec{v}=(\frac{t_2}{2},\frac{t_1}{2},t_{1d})$ along x-bond while $\vec{v}=(\frac{t_1}{2},\frac{t_2}{2},t_{1d})$ along y-bond. Here $c^{\dagger}_{i,A/B,\sigma}$ represents the operator creates an electron on site $i$ with sublattice $A/B$ and pseudospin $\sigma$.

In large $U$ limit, the spin model is then obtained as~\cite{DM1960} 
\begin{eqnarray}
H_{\rm eff}=J\sum_{\langle i,j \rangle}\vec{S}_i\cdot\vec{S}_j +\sum_{\langle i,j \rangle}\vec{D}_{ij}\cdot(\vec{S}_i \times \vec{S}_j)\; .
\label{eq:r2}
\end{eqnarray}
Here $J=\frac{4}{U}[(t_0)^2-\vec{v}\cdot\vec{v}]$ and $\vec{D}_{ij}=\frac{8\epsilon_i t_0 \vec{v}}{U}$ where $\epsilon_i$ is the change of sign in the adjacent bond~\cite{Krempa14,Carter2013}.

Note that when the bond retains the inversion symmetry, the DM vector $\vec{D}$ should vanish. However, due to the different rotation and tilting angles of oxygen octahedra between neighboring Ir atoms which break the inversion symmetry on the bond, the effective spin model of Eq.~\ref{eq:r2} is obtained.  The ground state of such spin Hamiltonian has a non-collinear form:
\begin{eqnarray}
m_{100}\sigma_x+m_{(010)}\sigma_y\tau_z+m_{(001)}\sigma_z\tau_z\; ,
\label{eq:r3}
\end{eqnarray}
where $m_{(010)}$ and $m_{(001)}$ represent sublattice antiferromagnetic orderings, while $m_{(100)}$ denotes a ferromagnetic component of ordering. The exact form and amplitudes of the magnetic orderings in Eq.~\ref{eq:r3} are related to the crystal symmetry and detailed hopping parameters on the bond. However, the specific magnetic pattern is not crucial to realize the QAH effect in single-layer iridates as long as TR symmetry is broken. 

In the absence of TR symmetry, the topological invariance characterizing the QAH effects is identified by the charge Chern number defined as,
\begin{equation}
C_p=\frac{1}{2\pi}\int d^2 {\bf{k}}\Omega^z_p({\bf{k}}),
\label{eq:a7}
\end{equation}
where $p$ is the band index and  $\Omega^z_p({\bf k})$ is z-component of p-th band Berry curvature ${\bf{\Omega}}_p({\bf{k}})$ given in the Appendix.
The quantized transverse Hall conductance $\sigma_{xy}$ is then given by
\begin{equation}
\sigma_{xy}=\frac{e^2}{h}\sum_{p \in occupied} C_p,
\label{eq:a8}
\end{equation}
where the sum goes over all occupied bands below Fermi energy $\epsilon_F$. 
For the single layer 2D Ir oxide, the quantized Hall conductivity is obtained as 
\begin{equation}
\sigma_{xy}=\frac{e^2}{h},
\label{eq:a9}
\end{equation}
indicating the topological invariance $C \equiv \sum_{p \in occupied }C_p=1$ related to the edge currents propagating along one direction on the sample boundary~\cite{Hatsugai93}
shown in Fig. ~\ref{fig:a1MTI}. 

Note that the QAH phase depends on the magnitude of the ordering. The different sizes of gaps at X and Y point appear after breaking b-glide symmetry see Fig.~\ref{fig:bglideno}. If the strength of the magnetic ordering reverses the bands at X point for instance, while keeping the gap at Y point intact, the system turns into the QAH phase with quantized $\sigma_{xy}$ of Eq.~\ref{eq:a9}. However, if the magnitude of the ordering is sufficiently large to reverse both bands at X and Y points, the system will thus turn to a trivial insulator. Thus, above the magnetic ordering temperature, the QAH phase should show up in a certain range of external magnetic field.


\section{Bilayer Iridates}

To realize the topological phases in the single layer IrO$_2$ layer, the b-glide symmetry should be externally broken.
This requires a strain field in a certain direction, which is not trivial in an experimental setting.
In this section, we propose two types of bilayer IrO$_2$ systems, which naturally hold topological phases without a lattice symmetry breaking perturbation.
Since the single IrO$_2$ layer has two different sets of rotation and tilting angles,
one way to engineer bilayer systems  is to stack two layers of $A$ and $B$ on top of each other. 
Note that  
$A$ and $B$ per unit cell have the rotation and tilting angle $(\theta,\phi)$ and $(-\theta,-\phi)$, respectively. 
Another way to stack two single layers is to make the second layer has different rotation and tilting set such as
$(\theta,-\phi)$ and $(-\theta,\phi)$ denoted by $C$ and $D$ sites, respectively.  
We call the first case ABAB stacking and the other case ABCD stacking: see Fig. ~\ref{fig:abab}.
The distance between top and bottom layers in both cases can be manipulated by the number of AMO$_3$ layers in between,
and the nature of topological phases is not altered by such quantitative changes.
Let us consider the ABAB stacking case first.

\subsection{ABAB stacking}

\begin{figure}[t]
\centering
\includegraphics[width=9cm]{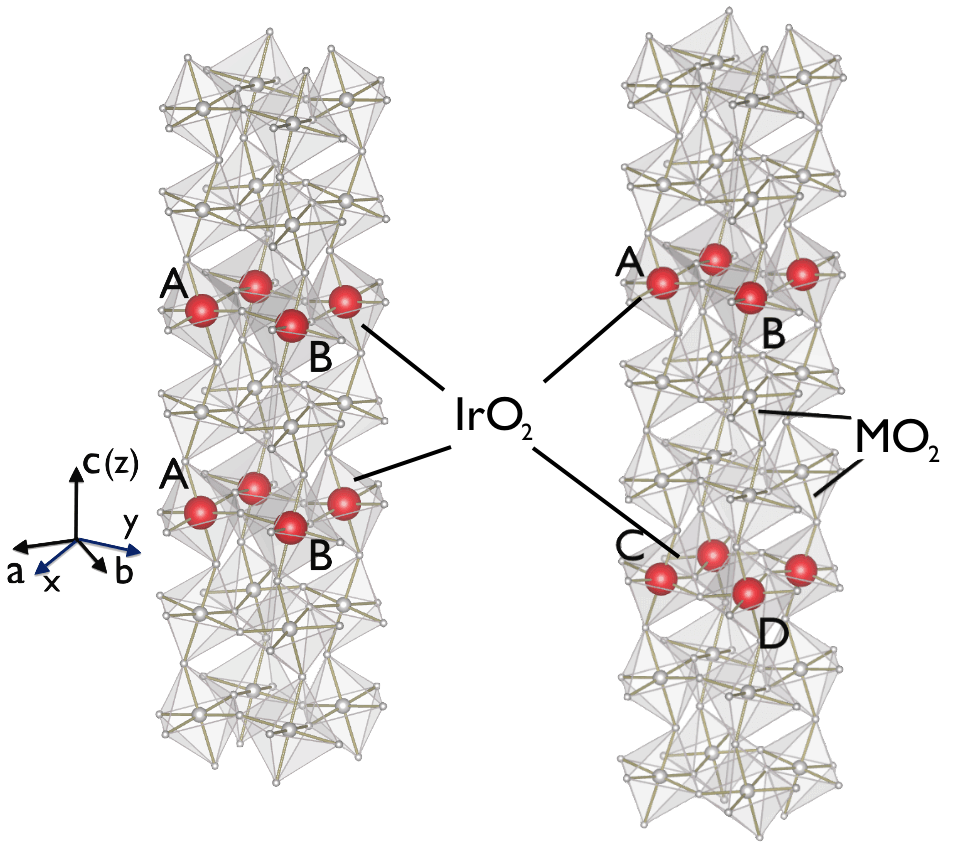}
\caption{(color online) (left) ABAB stacking with A=($\theta,\phi$) and B=($-\theta,-\phi$) types of octahedra rotation and tilting.
(right) ABCD bilayer stacking which contains A and B in the top layer, while C=$(\theta,-\phi)$ and D=$(-\theta,\phi)$ types of octahedra rotation and tilting in the bottom layer.}
\label{fig:abab}
\end{figure}

\begin{figure}[t]
\subfigure[$\phi=0$]{
\includegraphics[width=7cm]{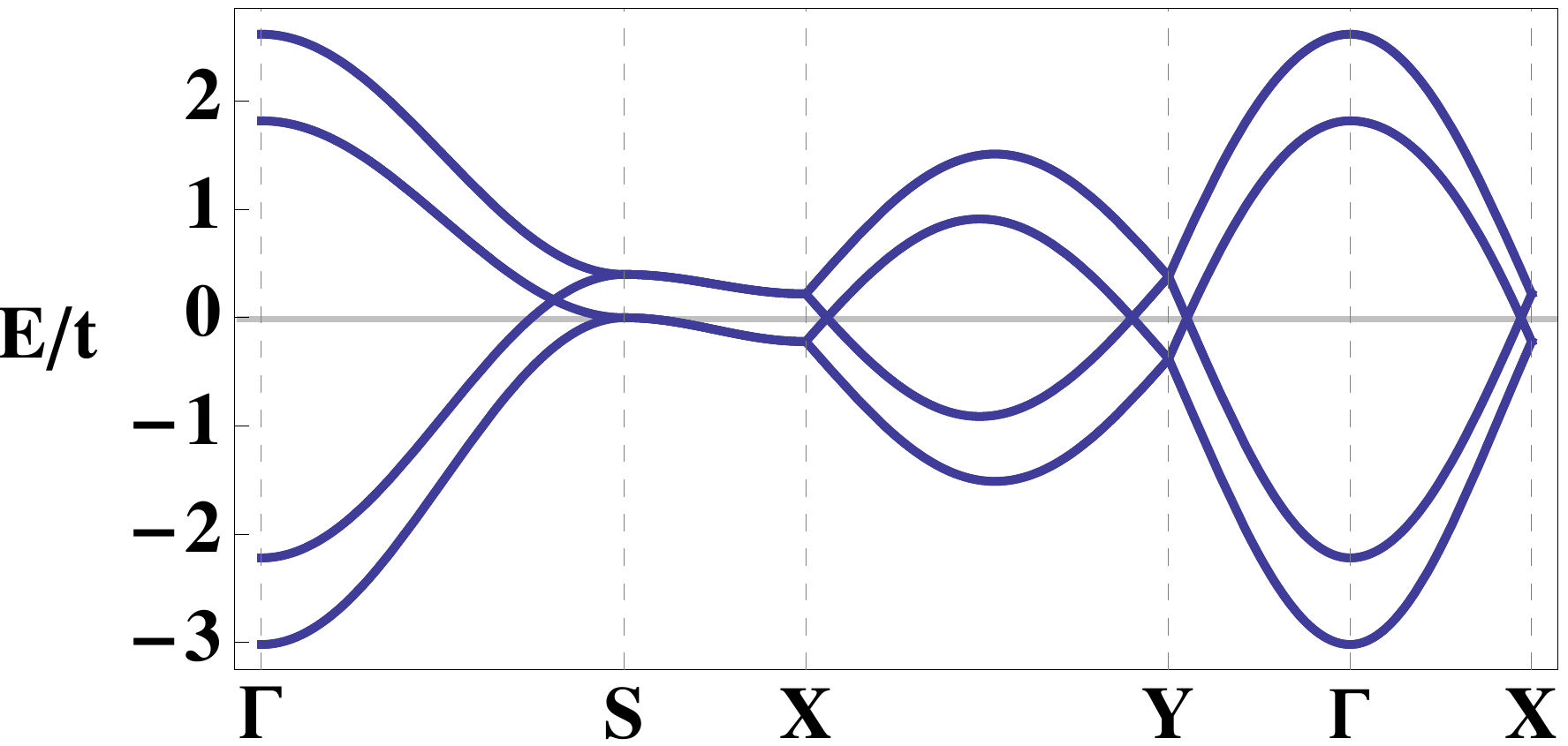}
\label{fig:bsnotilt}
}
\subfigure[Finite $\phi$ with ABAB stacking]{
\includegraphics[width=7cm]{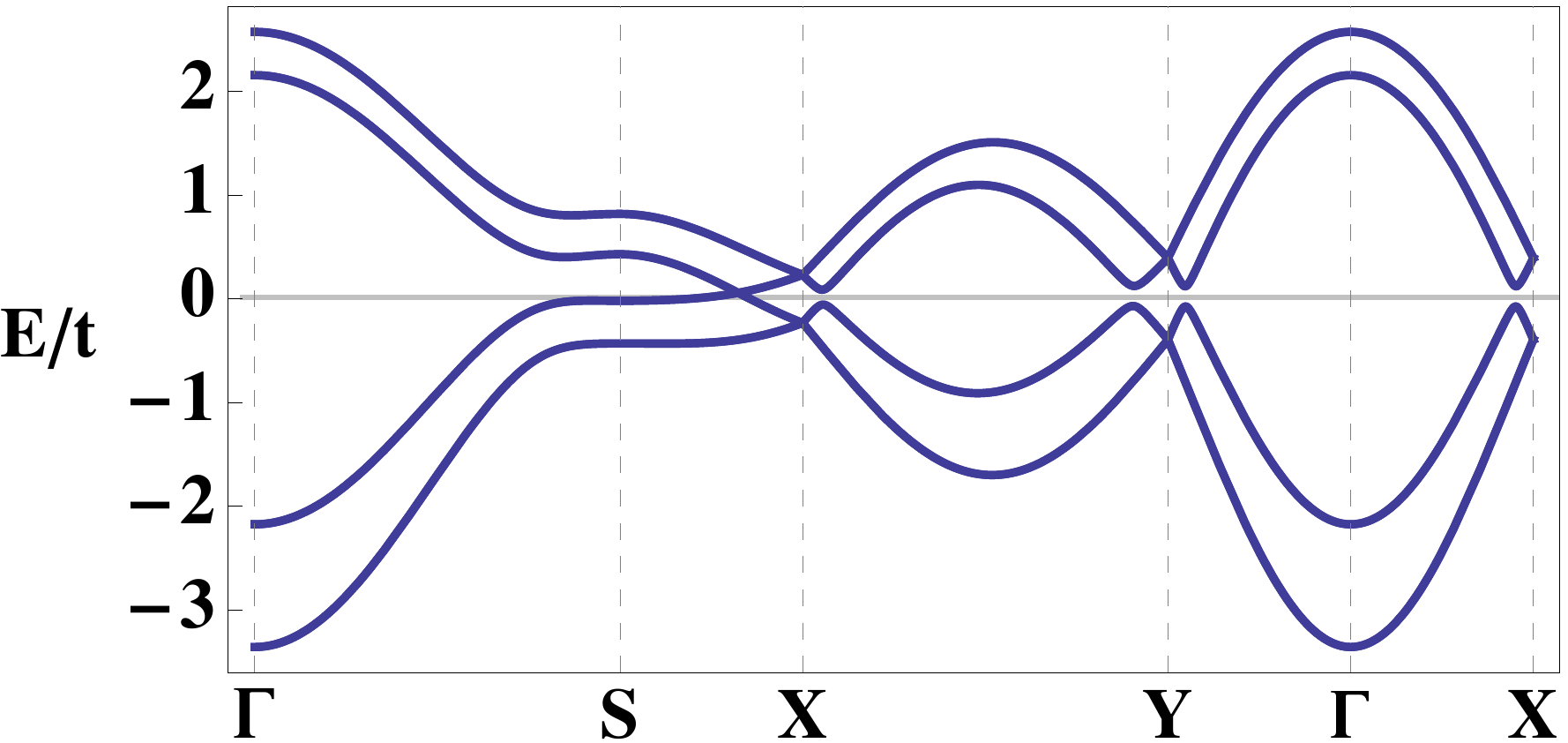}
\label{fig:bstilt}
}
\subfigure[Finite $\phi$ with ABCD stacking]{
\includegraphics[width=7cm]{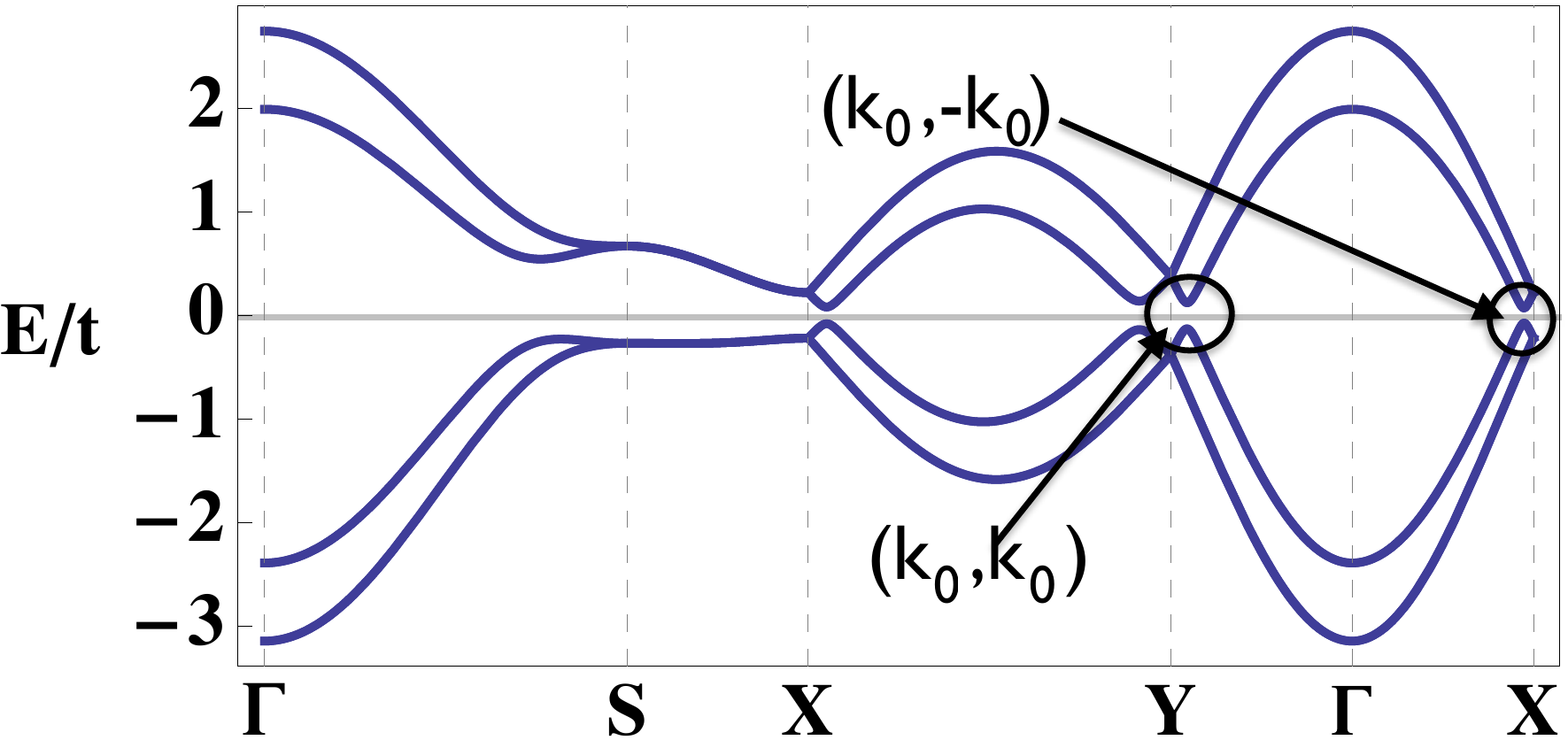}
\label{fig:abcdtilt}
}
\caption{(color online) Band structure for bilayer with no tilt effect ({\it i.e.} $\phi=0$) in octahedron environment (a) has a degenerate line circling $\Gamma$ point.~\cite{footnote1}(b)ABAB: Finite tilting lift the line node degeneracy but leaves one Dirac points protected by the b-glide symmetry along $S\rightarrow X$ for ABAB stacking. (c)Band structure for ABCD bilayer with finite tilting $\phi$. It has a band gap at $(k_0,\pm k_0)$ (circled out by band lines). The Fermi energy is $\epsilon_F=0$ indicated by gray solid lines.}
\label{fig:bsbilayer}
\end{figure}

\begin{figure*}[t]
\subfigure[FM]{
\includegraphics[width=7cm]{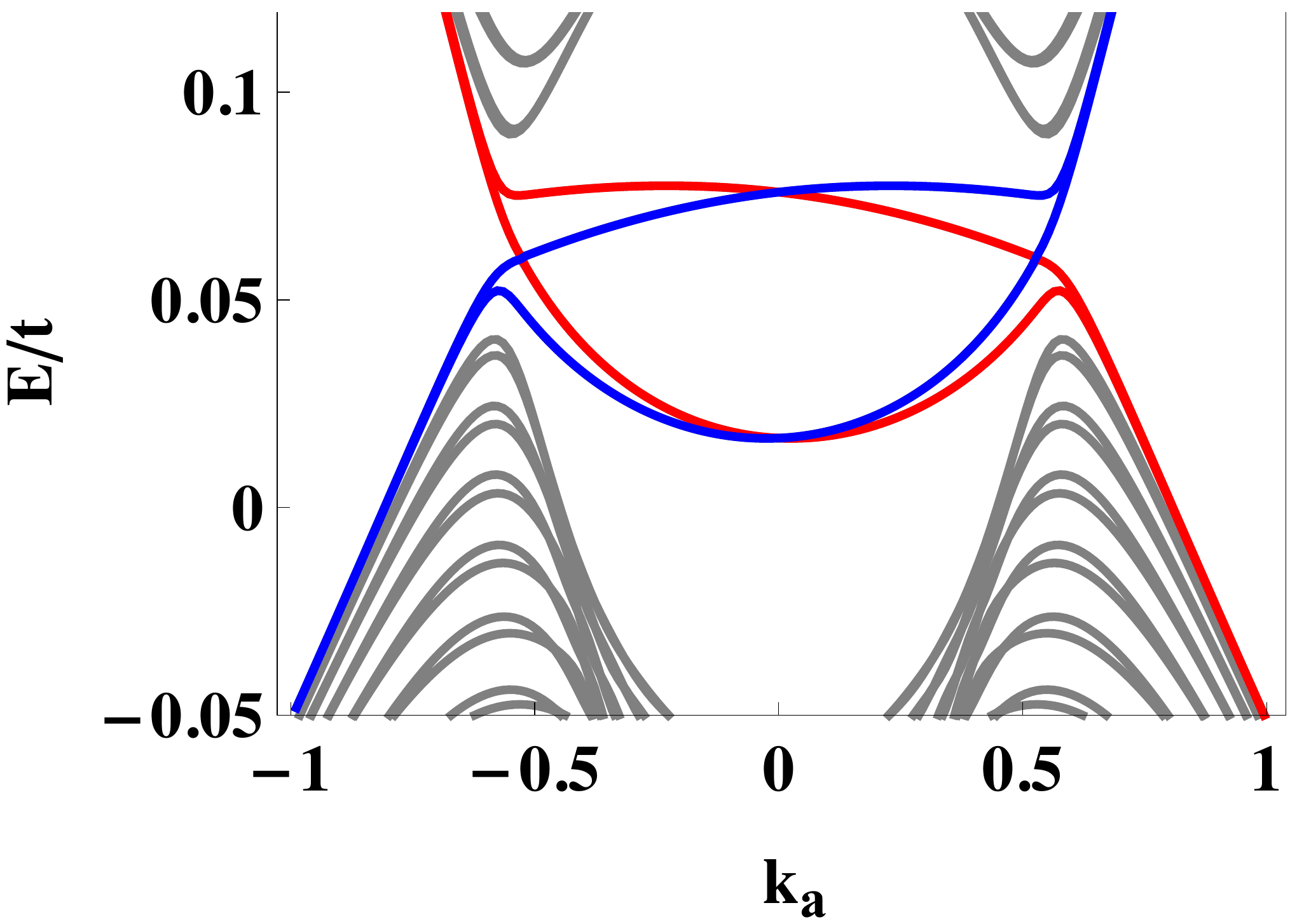}
\label{fig:bsfm}
}
\subfigure[AFM]{
\includegraphics[width=7cm]{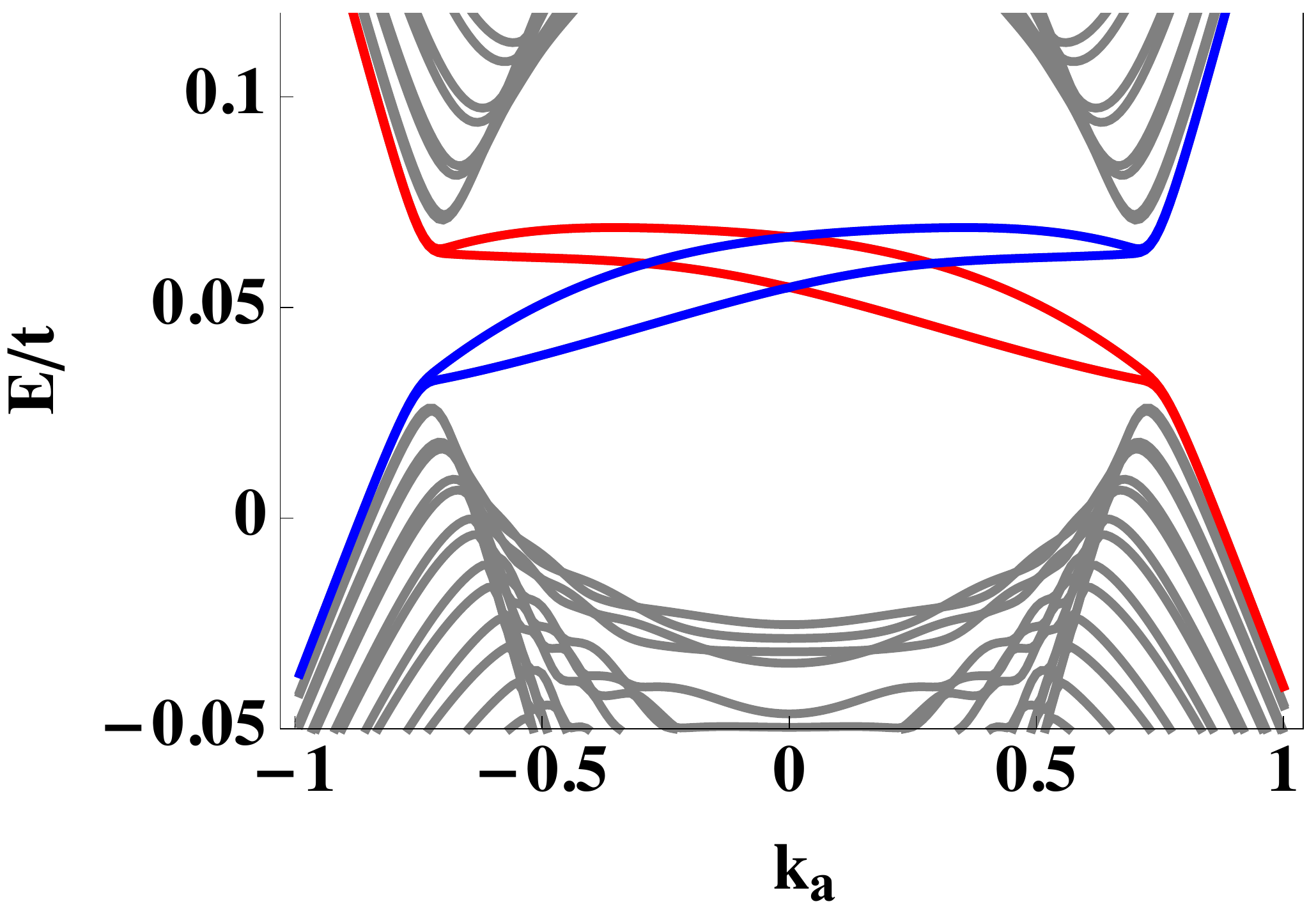}
\label{fig:bsafm}
}
\caption{(color online) Slab dispersion with (a) ferromagnetic (FM) with $m_{(110)}=0.09t$  and (b) antiferromagnetic ordering (AFM) with strength $m_{(1\bar{1}0)}=0.06t$.
Two gapless edge modes at $L=0$ and $L=N$ boundary are represented by red and blue, respectively. }
\label{fig:bs}
\end{figure*}

As presented in Fig.~\ref{fig:abab}, the ABAB bilayer structure with significant rotation and tilting can be obtained by inserting one layer band insulator material MO$_2$ (M=Zr, Hf) between two IrO$_2$ layers. The tight-binding Hamiltonian  is given by
\begin{equation}
H_{ABAB}({\bf{k}})=\sum_{i=1,2}  H^i_{0} ({\bf k})+H_{12}({\bf{k}}),
\label{eq:bs1}
\end{equation}
where $H^i_{0}$ represents a top ($i=1$) and bottom ($i=2$) IrO$_2$ layer and is same as Eq.~(\ref{eq:a3}).
$H_{12}$ contains the hopping terms between the two layers, and introducing another Pauli matrices $\vec{\nu}$ for the layer degree of freedom, it is written as
\begin{equation}
\begin{split}
H_{12}({\bf{k}})&=\epsilon_{di}({\bf{k}})\nu_x\\
&+\textrm{Re}(\epsilon_{dz}({\bf{k}}))\sigma_y\tau_y\nu_x+\textrm{Im}(\epsilon_{dz}({\bf{k}}))\sigma_z\tau_y \nu_y\\
&+\textrm{Re}(\epsilon_{z}({\bf{k}}))\sigma_y \tau_y \nu_x+\textrm{Im}(\epsilon_{z}({\bf{k}}))\sigma_x\tau_y \nu_x\\
&+\textrm{Re}(\epsilon_{z}^{\prime}({\bf{k}}))\sigma_y\tau_y \nu_y+\textrm{Im}(\epsilon_{z}^{\prime}({\bf{k}}))\sigma_x\tau_y\nu_y,
\end{split}
\label{eq:bs2}
\end{equation}
where
\begin{equation}
\begin{split}
\epsilon_{di}({\bf{k}})&=t_z+t_{(110)}\cos(k_x+k_y)+t_{(1\bar{1}0)}\cos(k_x-k_y),\\
\epsilon_{dz}({\bf{k}})&=t_{dz}(\cos(k_x)+\cos(k_y))+i t^{\prime}_{dz}(\sin(k_x)+\sin(k_y)),\\
\epsilon_{z}({\bf{k}})&=(t_{2z}\cos(k_y)+t_{1z}\cos(k_x))+i (k_x \leftrightarrow k_y),\\
\epsilon_{z}^{\prime}({\bf{k}})&=(t^{\prime}_{2z}\sin(k_y)+t^{\prime}_{1z}\sin(k_x))+i (k_x \leftrightarrow k_y).
\end{split}
\label{eq:bs3}
\end{equation}
Here 
$t_z$ is the NN hopping between two layers. $t_{(110)}$ and $t_{(1\bar{1}0)}$ are the third NN  intra-orbital hopping along $(110)$ and $(1\bar{1}0)$, respectively. $t_{dz}$ and $t^{\prime}_{dz}$ arise from $d_{yz}$ orbital to $d_{xz}$ orbital 
NNN hopping due to the rotation and tilting angles.  $t_{2z}, t_{1z}, t^{\prime}_{2z}$ and $t^{\prime}_{1z}$ are given by the overlap hopping integral between $d_{yz} (d_{xz})$ and $d_{xy}$-orbital. 
The parameters in tight-binding Hamiltonian Eq.~(\ref{eq:bs1}) are obtained based on Slater-Koster Method~\cite{Slater54} and  $(t_z,t_{(110)},t_{(1\bar{1}0)},t_{dz},t_{dz},t^{\prime}_{dz},t_{2z},t_{1z},t_{2z}^{\prime},t_{1z}^{\prime})/t=(-0.13,-0.01,-0.09,-0.03,-0.01,0.014,
0.01,0.062,0.01)$ for the same $\theta$ and $\phi$ used in the single layer.
The band structure in Fig.~\ref{fig:bsnotilt} shows that there are two line nodes around $X$ and $Y$ when $\phi=0$. However, a finite tilting $\phi$ lifts 
the band degeneracy, but keeps one pair of Dirac points along the high symmetry line $X \rightarrow S$ which is protected by the b-glide symmetry in Fig.~\ref{fig:bstilt}.

Due to the electronic correlation and DM interaction, a non-collinear magnetic ordering is expected. One example of non-collinear orderings has the form of 
\begin{equation}
m_{(110)} \left( \sigma_x + \sigma_y \right) + m_{(1\bar{1}0)}  \left( \sigma_x - \sigma_y \right) \tau_z+ m_{(001)}\sigma_z\tau_z,
\label{eq:bs4}
\end{equation}
Since an exact direction of magnetic ordering is not important for the topological nature,
we compute the Hall conductivity for (a) $m_{(110)} \neq 0$ and (b) $m_{(1\bar{1}0)} \neq 0$  cases.
For both cases, we found it is quantized as
\begin{equation}
\sigma^{bilayer}_{xy}=2\frac{e^2}{h},
\label{eq:bs5}
\end{equation}
which implies the charge Chern number defined in Eq.~(\ref{eq:a7}) for the entire valence bands  $C=2$.
The edge states computed in the zigzag slab geometry are shown
for (a) case in Fig.~\ref{fig:bsfm} and (b) case in Fig. ~\ref{fig:bsafm}, respectively.
This also confirms the existence of the two gapless edge modes propagating along the sample boundary.
Thus any magnetic ordering (or in-plane magnetic field) leads to a topological magnetic insulator with QAH effect in the 2D ABAB stacked bilayer Ir oxides.

The difference between the single layer and the bilayer ABAB stacking deserves some discussion, as the bilayer is obtained simply by stacking the AB single layer.
The Dirac nodes at X and Y TRIM points of the single layer are protected by the b-glide symmetry. However,
finite hopping integrals between two layers generate the different size of gaps at X and Y points in the ABAB bilayer system, and
the Dirac point is shifted to a non-symmetric point. Thus any magnetic field or magnetic ordering that breaks the b-glide symmetry 
would turn the system into a topological magnetic insulator.  On the other hand in the single layer, a magnetic field and/or ordering that breaks TR and the b-glide symmetry simultaneously induces the same
strength of gap at the X and Y points
making the system a trivial insulator. Thus an external b-glide symmetry breaking perturbation is necessary to generate different gaps at $X$ and $Y$ in order to realize QAH insulator in 
the single layer case. Below we consider the other type of layer stacking, which offers various topological phases.

\subsection{ABCD stacking}

\begin{figure*}[t]
\begin{minipage}[h]{0.3\textwidth}
\subfigure[QAH]{
\includegraphics[width=5.3cm]{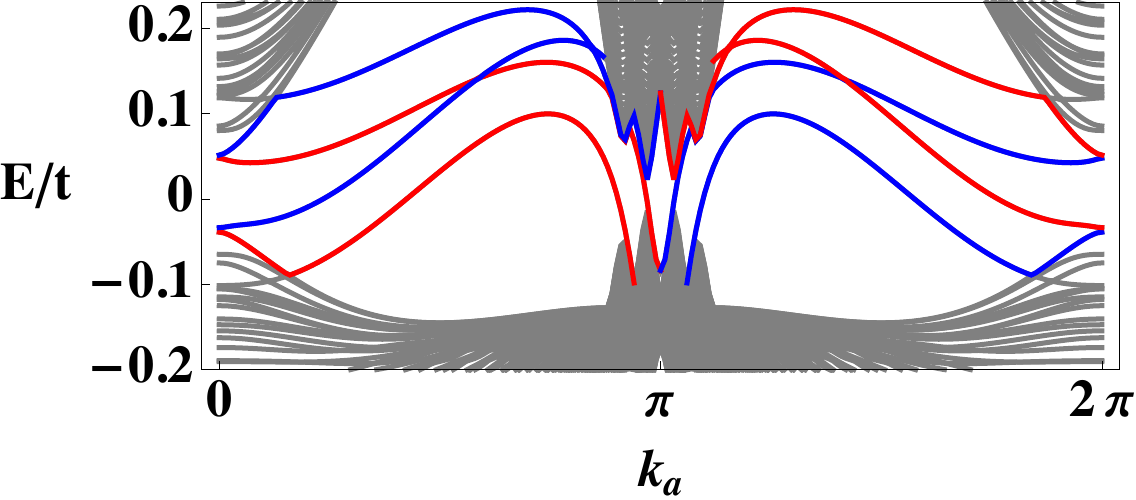}
\label{fig:mvh}
}
\subfigure[QVH]{
\includegraphics[width=5cm]{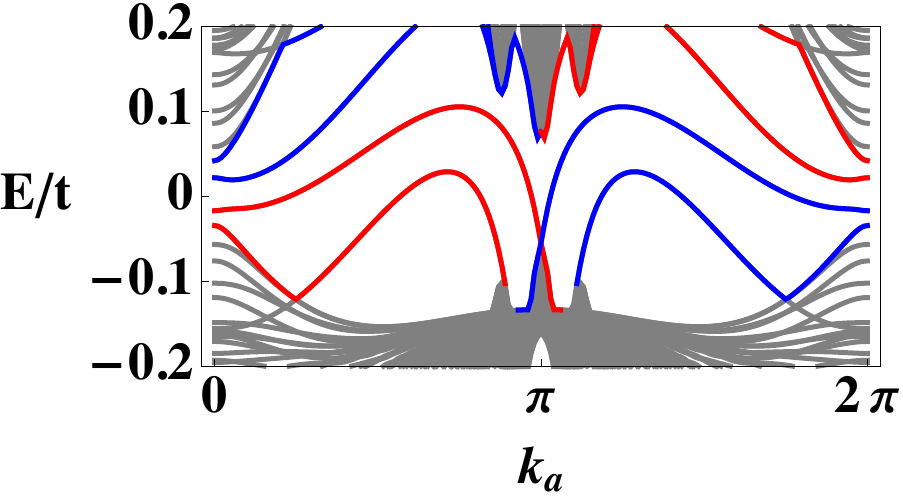}
\label{fig:tci}
}
\end{minipage}
\begin{minipage}[h]{0.3\textwidth}
\subfigure[Phase Diagram]{
\includegraphics[width=6.cm]{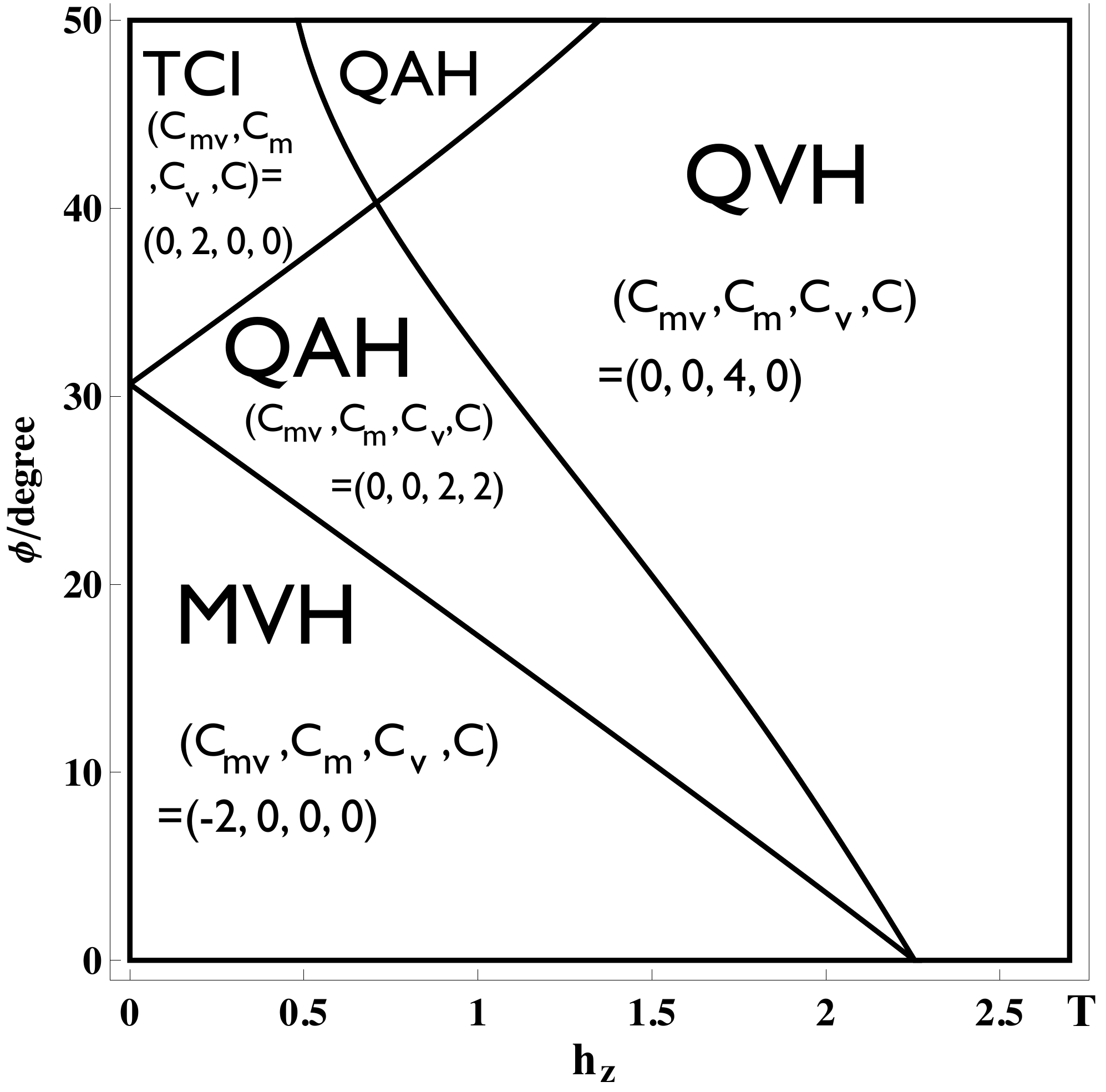}
\label{fig:pd}
}
\end{minipage}
\begin{minipage}[h]{0.37\textwidth}
\subfigure[MVH]{
\includegraphics[width=5.2cm]{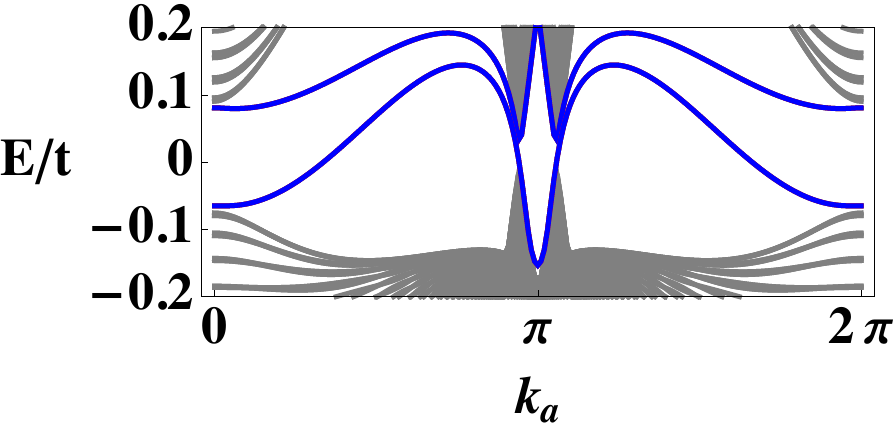}
\label{fig:mvh}
}
\subfigure[TCI]{
\includegraphics[width=5cm]{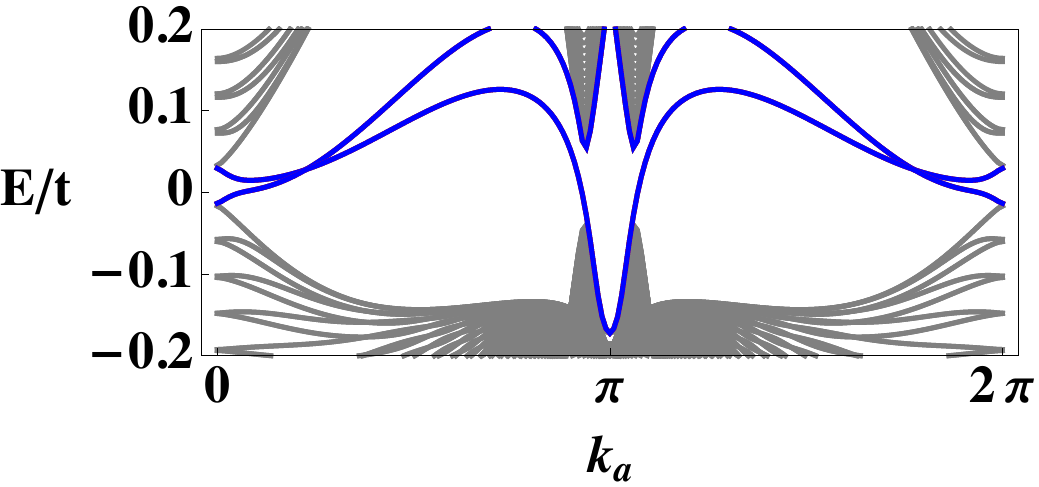}
\label{fig:tci}
}
\end{minipage}
\caption{(color online) Phase diagram when rotation degree is $\theta=13^{\circ}$ in the middle panel (c) plotted as z-direction exchange field $h_z$ in the unite of Tesla (T) versus tilting degree $\phi$. Different phases has been characterized by different topological invariants $(C_{mv},C_m,C_v,C)$. The edge state for each phase has been displayed in (a) QAH, (b) quantized valley Hall (QVH), (d) mirror valley Hall (MVH) and (e)topological crystalline insulator (TCI). Two gapless edge modes in (a), (b), (d) and (e) at $L=0$ and $L=N$ boundary are represented by red and blue, respectively. Edge states are purple (mixed color of red and blue) in (d) and (e) because of the degeneracy between edge modes at $L=0$ and $L=N$. See the main text for finite $C_{mv}$ and $C_m$ related to these edge modes.}
\label{fig:pdedge}
\end{figure*}

The crystal structure with ABCD stacking is displayed in Fig.~\ref{fig:abab}.  
The tight-binding Hamiltonian for this stacking is given by 
\begin{eqnarray}
H_{ABCD}({\bf{k}})=\sum_{i=\pm} H^i_0({\bf{k}})+H^\prime_{12}({\bf{k}}),
\label{eq:b1}
\end{eqnarray}
where 
\begin{eqnarray}
H^{\pm}_{0}({\bf{k}})&=&\epsilon^\prime({\bf{k}}){\bf{I}}+\epsilon_0 ({\bf{k}})\tau_x+\epsilon_{1d}({\bf{k}})\sigma_z\tau_y\nonumber\\
&&\pm (\epsilon_y({\bf{k}})\sigma_y\tau_y+\epsilon_x({\bf{k}})\sigma_x\tau_y),\\
H^\prime_{12}({\bf{k}})&=&\epsilon_{di}({\bf{k}})\nu_x+\epsilon_{12}({\bf{k}})\tau_x\nu_x
+t_z^{\prime}(\sigma_y+\sigma_x)\tau_z\nu_y.\nonumber
\label{eq:b2}
\end{eqnarray}
The various dispersions $\epsilon({\bf k})$s  in $H^{\pm}_{0}$ have the same expression as Eq.~(\ref{eq:a4}), which represent intra-layer hopping integrals for top ($i=+$) and bottom ($i=-$) layer.
$H^\prime_{12}$ contains hopping paths between the two layers, and the dispersion $\epsilon_{di}({\bf k})$ is the same as Eq.~(\ref{eq:bs3}). $t_z^{\prime}$ 
represents the 1D orbital to $d_{xy}$-orbital hopping between the layers,  and
\begin{eqnarray}
\epsilon_{12}({\bf{k}})=t_{12}(\cos(k_x)+\cos(k_y)),
\label{eq:b3}
\end{eqnarray}
where $t_{12}$ denotes the NNN inter-layer intra-orbital hopping. 

In addition to the b-glide symmetry $\hat{\Pi}_b$ in Eq.~(\ref{eq:a1}), 
there exists another glide plane which transfers between top and bottom layers in this bilayer system. 
\begin{eqnarray}
\hat{\Pi}_{layer}=\frac{i}{\sqrt{2}}(\sigma_x+\sigma_y)\tau_x\nu_x \hat{k}_{layer},
\label{eq:b4}
\end{eqnarray}
where $\hat{k}_{layer}$ is the operator that interchanges $k_x$ with $k_y$ as $\hat{k}_{layer}: (k_x,k_y) \rightarrow (-k_y, -k_x)$. By computing the commutator of $\hat{\Pi}_{layer}$ with $H_{ABCD}({\bf{k}})$, it is straightforward to verify that $[\hat{\Pi}_{layer},H_{ABCD}]$=0. 

The band dispersion is shown in Fig. ~\ref{fig:abcdtilt}. The set of  tight-binding parameters is given by $(t_z,t_{(110)},t_{1\bar{1}0},t_{12},t_z^{\prime})/t=(-0.23,-0.01,-0.09,-0.11,-0.04)$ for the same $\theta$ and $\phi$ in the single layer. The hopping amplitude changes as a function of distance and has been estimated by introducing a scaling function $1/r^5$. There are two line nodes appear when the tilting degree vanishes \textit{i.e.} $\phi=0$ and those degeneracies are gapped out after introducing some finite tilting as shown in Fig.~\ref{fig:abcdtilt}

To analyze the topological nature of the bilayer system, we introduce the combined symmetry of $\hat{\Pi}_b$ and $\hat{\Pi}_{layer}$ such that
$\hat{\Pi}_{mirror} \equiv \hat{\Pi}_b \hat{\Pi}_{layer}=i\sigma_z\nu_x \hat{k}$ with $\hat{k}$: $(k_x,k_y)\rightarrow(-k_x,-k_y)$. 
Since the Hamiltonian is even under $\hat{k}$,  $[i\sigma_z\nu_x,H_{ABCD}]=0$. Furthermore, the low energy effective Hamiltonian can be brought into a block diagonalized form
 near $X$ and $Y$ TRIM points with each block labeled by the eigenvalues of $\sigma_z\nu_x$, given by
\begin{eqnarray}
H^{{\rm eff}}_{\pm,X/Y}=\vec{A}_{\pm,X/Y}({\bf{k}})\cdot \vec{\sigma},
\label{eq:b5}
\end{eqnarray}
where $\pm$ subscripts are assigned to reflect the eigenvalues of the combined operator $\hat{\Pi}_{mirror}$.
The explicit expression of vector $\vec{A}_{\pm,X/Y}({\bf{k}})$ is presented in the Appendix. 
One way to glimpse the novel topological phases lying behind the gapped band structure is to evaluate the topological charges~\cite{Ezawa12}
defined by the mirror valley (MV) Chern number $C_{mv}$, valley Chern number $C_v$, and mirror Chern number $C_m$ in addition to the charge Chern number $C$ at $X$ and $Y$ TRIM points: 
\begin{eqnarray}
C_{mv}&=& \frac{1}{2} (C_{+,X}-C_{-,X}+C_{-,Y}-C_{+,Y}),\nonumber\\
C_m&=&\frac{1}{2}(C_{+,Y}-C_{-,Y}-C_{-,X}+C_{+,X}),\nonumber\\
C_v&=& (C_{+,X}+C_{-,X}-C_{+,Y}-C_{-,Y}),\nonumber\\
C&=& ( C_{+,X}+C_{-,X}+C_{+,Y}+C_{-,Y}).
\label{eq:b7}
\end{eqnarray}
The charge Chern number $C$ is sum of all Chern number $C_{\pm,X/Y}$ associated with valleys ($X/Y$) and mirror symmetry eigenvalues ($\pm$). The valley-Chern/mirror-Chern number $C_v$/$C_m$ is odd only under the interchange of two valleys/mirror symmetry eigenvalues.  The mirror-valley-Chern number $C_{mv}$, however, is odd under the interchange of valleys and mirror symmetry eigenvalues, respectively. The computation details of $(C_{mv},C_m,C_v,C)$ and the explicit expressions are presented in the Appendix.

A phase diagram contains various phases~\cite{footnote}
including mirror valley Hall phase, topological crystalline insulator phase, 
QAH phase and quantized valley Hall phase with distinguished topological features, as displayed in Fig.~\ref{fig:pd}. The phases listed here are robust against disorder as long as it preserves the symmetry associated with each phase~\cite{KanePRL05-2,Hsieh12,ZhangPRL11}.
The vertical axis is the degree of tilting angle $\phi$ and 
the horizontal axis corresponds to the strength of $z$-component of the magnetic exchange field and/or ordering. 
The phase boundaries can be modified depending on the magnetic ordering
or exchange field pattern,  but the qualitative picture of the phase diagram is not sensitive to the choice of magnetic ordering direction, as long as there is a finite z-component of ferromagnetic $h_z$ or antiferromagnetic ordering of $m_z$.
Thus we only tune the strength of $h_z$ for simplicity.  In Fig. ~\ref{fig:pd}, $h_z$ is estimated in Tesla using the tight binding parameters discussed above,
and set $t \sim 100 meV$. 

Each phase separated by thick black line in Fig.~\ref{fig:pd} is charactered by the unique set of topological invariance $(C_{mv},C_m,C_v,C)$ defined in Eq.~(\ref{eq:b7}).  The edge states shown in Fig. 7(a), 7(b), 7(d) and 7(e) 
 are obtained with the slab geometry under the same boundary condition with ABAB stacking case described in the last section.

 
The bilayer with small tilting angle is characterized by mirror valley Hall phase with $C_{mv} = -2$.
The valley physics in mirror valley Hall phase manifests explicitly in the edge state dispersion in Fig.~\ref{fig:mvh}.  
When the degree of tilting angle $\phi$ increases, it becomes a topological crystalline insulator with $C_m =2$.
The large tilting degree is able to inverse the sign of one of the mass term near $X$ or $Y$, and thus modifies the topology of the system. 
The edge state dispersion for topological crystalline insulator phase in Fig.~\ref{fig:tci} has two pairs of gapless currents moving along opposite directions on each boundary. Each pair of edge modes carries opposite mirror eigenvalues. As the name suggest, these two pairs of gapless edge states are indeed protected by $\hat{\Pi}_{mirror}$.
A TR breaking term will not lift the degeneracy between edge states as long as the perturbation preserves $\hat{\Pi}_{mirror}$.


By tuning the strength of $h_z$, the QAH phase arises. In the QAH phase, two gapless edge states localized at $L=0$ propagate along the same direction.
Each one contributes $e^2/h$ to the Hall conductance and the total Hall conductivity, when Fermi energy has been tuned inside the bulk gap is given by
\begin{equation}
\sigma_{xy}=2\frac{e^2}{h}.
\label{eq:b12}
\end{equation}
However, in quantized valley Hall phase, within valley $X$ ($Y$), the two edge states localized at $L=0$ propagating along the same direction lead to quantized vally-Hall conductivity $\sigma_{xy}^v$.
\begin{equation}
\sigma_{xy}^v=C_v\frac{e^2}{h}=4\frac{e^2}{h}.
\label{eq:b11a}
\end{equation}
In order to detect the anomalous Hall conductivity $\sigma_{xy}^v$, photon illumination with circularly polarized light can be used which has been reported in the monolayer MoS$_2$ transistors~\cite{Mak2014}.  Since these two valleys are related by the inversion symmetry, it requires to break the inversion symmetry to measure the valley-Hall conductance in Eq.~\ref{eq:b11a}.

The mirror and mirror valley Chern numbers $(C_m, C_{mv})$ can be understood through the behavior of edge modes localized at $L=0$ for instance. When the system is in mirror valley Hall phase, there are four edge modes at $L=0$ or $L=N$ as shown in Fig.~\ref{fig:mvh}. Two edge modes are propagating from left to right labeled with $(-,X)$ and $(+,Y)$, respectively. The other two are flowing along the opposite direction named as $(+,X)$ and $(-,Y)$, respectively. Here $(\pm,X/Y)$ means the edge state carries $\pm$ quantum number which is the eigenvalue of $\sigma_z\nu_x$ and the valley degree of freedom $X/Y$. Thus $C_{mv}$ is finite. When the gap is reversed at $X$, the propagating direction of the edge modes $(\pm,X)$ will reverse and result in a non-vanishing $C_m$. Therefore, the system is a topological crystalline insulator as shown in Fig.~\ref{fig:tci}. ARPES has proven
to be ideally suited to detect topological signatures of TCIs~\cite{Tanaka12}; such
methods can be in principle generalized to detect the MVH insulator.

As we emphasize above, a finite bilayer hopping integral is crucial to achieve the QAH phase when TR symmetry is broken,
because the z-axis ferromagnetic exchange field $h_z \sigma_z$ (or sublattice antiferromagnetic ordering $m_z \tau_z \sigma_z$) 
has to overcome $t_z$ to reverse the sign of Berry curvature around $X$ or $Y$ in order to enter the QAH insulator phase (see the Appendix for the proof).
Using the current tight binding parameters, the strength of $h_z$ needs to be about a few Tesla as shown in Fig. \ref{fig:pd}.
Since the critical strength of $h_z$ is tuned by the strength of $t_z$,
it is desirable to make the bilayer hopping $t_z$ smaller, which can be controlled by the spacing
between the layers as shown in of Fig.~\ref{fig:abab}.

\section{Conclusions}

A recent experiment has reported successful growth of  Ir oxide superlattice [(SrIrO$_3$)$_n$, SrTiO$_3$] with controllable number of layers $n$,
  which tailors a spin-orbit magnetic insulator for $n=1$ and $2$. ~\cite{Matsuno14} 
Due to the smaller lattice constant in TiO$_2$ compared with IrO$_2$, it was expected that there are alternating rotations of Ir octahedra, but lacking the tilting ($\phi$) of octahedra to keep the tetragonal crystal structure of SrTiO$_3$. This was confirmed by the magnetic ordering patterns in $n=1$ and 2 superlattices, consistent with the first principle
calculations. ~\cite{Matsuno14} 
However, topological phases have not been observed in these superlattices, even though bulk SrIrO$_3$ orthorhombic perovskites possess a crystal-symmetry-protected nodal line.~\cite{CarterPRB12} 

One essential ingredient to realize any topological insulator is a Rashba-like SOC.  In the J$_{\rm eff}$=1/2 wavefunction formed by a strong atomic SOC, this Rashba-like SOC 
is generated by finite hopping integrals between different $J_z = \pm 1/2$ states. For example,
finite hopping paths between $d_{xy}$ and $d_{xz/yz}$ generate Rashba-like SOC terms in $J_{\rm eff}=1/2$ basis since $d_{xy}$ up-spin and one-dimensional orbitals of $d_{xz/yz}$ up-spin belong to different $J_z$ states.  In layered perovskite systems, this is possible when the hopping path does not
respect the mirror symmetry under $z \rightarrow -z$, as $d_{xy}$ is even while $d_{xz/yz}$ is odd under this operation.
Thus the alternating octahedra rotations and tiltings are necessary for topological phases in layered perovskites.

We propose topological phases in Ir oxide superlattices or films. 
Different topological phases were found depending on how the TR and crystal symmetries are broken. We consider three types of superlattice: single layer, bilayer with ABAB stacking and bilayer with ABCD stacking. A brief summary of our results is listed below. 

For the single-layer Ir oxide,  the Dirac dispersion at X and Y TRIM points is protected by the b-glide symmetry. 
When this b-glide symmetry is broken for instance by an uniaxial pressure, it reveals a 2D topological insulator by gapping the Dirac nodes. 
In the presence of a magnetic ordering or external magnetic field, the system becomes a topological magnetic insulator with QAH effects.

In the bilayer Ir oxides, we consider two different types of stacking. (1) For ABAB stacking, the system is a semimetal with two nodal points at $\epsilon_F$. Any finite magnetic field for any direction expect $[1\bar{1}0]$ axis or magnetic ordering turns the system into a topological magnetic insulator with QAH effects. Thus, the topological magnetic insulator in ABAB stacking is more realizable in current experiment setting than the single layer case.  (2) In the ABCD stacking case, due to an additional mirror symmetry $\Pi_{mirror}$, it provides a richer phase diagram. Besides the QAH phase, there are two additional phases: TCI with non-trivial mirror Chern number and MVH insulators with quantized mirror-valley Chern number.

Experimentally, these superlattices or films are grown along the [001] axis, which can be achieved by a most standard PLD growing technique. 
To test the proposal,  ARPES measurement can be  employed to
investigate the Dirac points in these superlattices when TR symmetry is preserved,
and Hall conductivity measurement should exhibit the QAH effect when a magnetic ordering occurs or an external magnetic field is applied.


\begin{acknowledgments}
This work was supported by the NSERC of Canada and the center for Quantum Materials at the University of Toronto.
\end{acknowledgments}

\appendix*
\section{Analytical Results of ABCD Bilayer}

Applying the following canonical transformation in $\sigma$ and $\nu$ space, 
\begin{eqnarray}
\sigma_{\pm}&\rightarrow&\sigma_{\pm} \nu_z,\nonumber\\
\nu_{\pm}&\rightarrow&\nu_{\pm}\sigma_z,
\label{eq:A5}
\end{eqnarray}
the Hamiltonian in Eq.~(\ref{eq:b1}) can be brought it into a block diagonalized form.
\begin{equation}
H^{\prime}=
\begin{pmatrix}
H^{\prime}_+ & 0 \\
0 & H^{\prime}_-
\end{pmatrix},
\label{eq:A6}
\end{equation}
with 
\begin{eqnarray}
H^{\prime}_{\pm}&=&\pm\epsilon_{di}({\bf{k}})\sigma_z+\epsilon_0({\bf{k}})\tau_x+\epsilon_y({\bf{k}})\sigma_y\tau_y\nonumber\\
&+&\epsilon_x({\bf{k}})\sigma_x\tau_y \pm t_z^{\prime}(\sigma_x-\sigma_y)\tau_z,
\label{eq:A6}
\end{eqnarray}
where $\pm$ subscripts are assigned to reflect the eigenvalues of $\sigma_z\nu_x$ and the basis we choose for $H^{\prime}$ is a set of the eigenvectors of $\sigma_z\nu_x$.

Let us consider the upper block Hamiltonian $H^{\prime}_+$ near $X$ point for now. By computing the eigenvalues of $H^{\prime}_+$ in Eq.~(\ref{eq:A6}) along $X \rightarrow \Gamma$, the location where the band gap vanishes near $X$ is given by
\begin{eqnarray}
\cos(k_{0}^{\pm})=\pm \frac{t_z^{\prime}}{t_1-t_2}.
\label{eq:A9}
\end{eqnarray}
The two solutions $(k_0^{\pm},- k_0^{\pm})$ in Eq.~(\ref{eq:A9}) in fact has the same topological properties. For convenience, only one point $(k_0^+, -k_0^+)\equiv (k_0,-k_0)$ will be taken into account.

Effective two band Hamiltonian can be obtained by projecting the $H^{\prime}_+$ to the relevant two bands $|\phi\rangle$ and $|\varphi \rangle$ at $(k_0,-k_0)$. 
\begin{eqnarray}
|\phi \rangle &=&\frac{1}{\sqrt{2}}(-|1,\downarrow \rangle+|2,\downarrow \rangle),\nonumber\\
|\varphi \rangle &=&\frac{1}{\sqrt{2}}(|1,\uparrow \rangle+|2,\uparrow \rangle),
\label{eq:A10}
\end{eqnarray}
where 1(2) refers to top (bottom) layer and $\uparrow$($\downarrow$) for $|J_z=\frac{1}{2}(-\frac{1}{2})\rangle$. Follow the perturbation theory, the effective two band Hamiltonian around $X$ is written as
\begin{equation}
H^{\textrm{eff}}_{+,X}=\hat{P}_{0}H^{\prime}_+\hat{P}_{0}=\vec{A}_{+,X}({\bf{k}})\cdot \vec{\sigma},
\label{eq:A11}
\end{equation}
where the projecting operator is $\hat{P}_{0}=|\phi\rangle \langle \phi |+|\varphi\rangle \langle \varphi |$ and each component of ${\bf A}_{+,X}$ is given by
\begin{equation}
\begin{split}
A^z_{+,X}&=t_z+t_{(110)}+t_{(1\bar{1}0)}^{\prime}+ t_0^{{\prime}}\equiv \delta_{X},\\
A^{y/x}_{+,X}({\bf{k}})&=t_{1}^{\prime} \delta k_{y/x}- t_{2}^{\prime} \delta k_{x/y}. 
\end{split}
\label{eq:A12}
\end{equation}
Here $t_{1}^{\prime}=t_{1} \sin(k_{0})$, $t_{2}^{\prime}=t_2 \sin(k_{0})$,$t^{{\prime}}_{(1\bar{1}0)}=t_{(1\bar{1}0)}\sin(k_0)$, $t_0^{\prime}=4t_0 \cos(k_0)$ and  $\delta k_x\equiv k_x-k_0, \delta k_y \equiv k_y+k_0$ for the following calculation. The Berry curvature for p-th band is given as $
{\bf{\Omega}}_p ({\bf{k}})={\bf{\nabla}}_k \times (i \langle p, {\bf{k}}| {\bf{\nabla}}_k |p, {\bf{k}}\rangle)$. Thus the Berry curvature for the lowest band of $H_{+,X}^{{\rm eff}}$ in Eq.~(\ref{eq:A11}) is
\begin{equation}
\Omega^z_{+,X}({\bf{k}})=\frac{((t^{\prime}_{2})^2-(t^{\prime}_{1})^2) \delta_{X}}{|\vec{A}_{+,X}|^3}.
\label{eq:A13}
\end{equation}

The Chern number can be computed using the formula Eq.~(\ref{eq:a7}) given in the main text and the expression is quite straightforward.
\begin{equation}
C_{+,X}={\rm sign} (\delta_{X}).
\label{eq:A14}
\end{equation}

Following the same procedure for lower block $H^{\prime}_-$ around $X$ and $\pm$ block around $Y$, the Chern number is given by
\begin{eqnarray}
C_{\pm,X}&=&\pm{\rm sign} (\delta_{X}),\nonumber\\
C_{\pm,Y}&=&\mp{\rm sign} (\delta_{Y}),
\label{eq:A15}
\end{eqnarray}
where $\delta_Y=t_z+t^{\prime}_{(110)}+t_{(1\bar{1}0)}+ t_0^{{\prime}}$ with $t^{\prime}_{(110)}=t_{(110)}\sin(k_{0})$.

Various topological charges in bilayer system has been identified in Eq.~(\ref{eq:b7}). Plug in the expression of Eq.~(\ref{eq:A15}), we have
\begin{eqnarray}
C_{mv}&=&{\rm sign}(\delta_X)+{\rm sign}(\delta_Y),\nonumber\\
C_m&=&{\rm sign}(\delta_X)-{\rm sign}(\delta_Y),\nonumber\\
C_v&=&C=0.\nonumber\\
\label{eq:A17}
\end{eqnarray}

Here by considering the magnetic field along z-direction $h_z\sigma_z$ with $h_z>0$, TR can be explicitly broken.   
The only modification in two band effective Hamiltonian is the mass term appears in Eq.~(\ref{eq:A13}) which changes to 
\begin{equation}
\delta_{X/Y} \rightarrow h_z\pm \delta_{X/Y}.
\label{eq:A21}
\end{equation}
And the Chern numbers in Eq.~(\ref{eq:A15}) have the following expressions. 
\begin{equation}
\begin{split}
C_{\pm,X}&=\textrm{sign}(h_z\pm \delta_{X}),\\
C_{\pm,Y}&=-\textrm{sign}(h_z\pm \delta_{Y}).
\end{split}
\label{eq:A22}
\end{equation}
The explicit expression of $C_{mv}$, $C_{m}$ $C_v$ and $C$ in Eq.~(\ref{eq:b7}) can be modified accordingly based on Eq.~(\ref{eq:A22}). The above analysis indicates the phase transition is driven by the z-direction magnetic field $h_z$.



\begin{thebibliography}{99}
\bibitem{footnoteaxis}The spin quantization axis was chosen to match the local z-axis of oxygen octahedron.
\bibitem{footnote1}The nodal points shown in Fig.~\ref{fig:bsnotilt} are connected and form a close degenerate line with dispersion $\cos(k_x)+\cos(k_y)=\frac{2t_z}{\sqrt{(2t_0-t_{12})^2+(t_{1d})^2}+\sqrt{(2t_0+t_{12})^2+(t_{1d})^2}}$; tight-binding parameters in the dispersion is given in the main text.
\bibitem{footnote}Very small amount of tilting degree may lead to electron and hole pockets of Fermi surface. But it does not change the topological nature of the system qualitatively.
\end{thebibliography}
\end{document}